\documentclass[10pt,twocolumn,showpacs,preprintnumbers,amsmath,amssymb,aps,prc,superscriptaddress]{revtex4-1}
\usepackage{lineno}
\usepackage{graphicx}% Include figure files
\usepackage{dcolumn}% Align table columns on decimal point
\usepackage{bm}% bold math

\begin{document}

%\preprint{Version 1.31}

\title{Beam-Energy Dependence of Charge Balance Functions from Au+Au Collisions at RHIC}% Force line breaks with \\\vec{}

\affiliation{AGH University of Science and Technology, Cracow 30-059, Poland}
\affiliation{Argonne National Laboratory, Argonne, Illinois 60439, USA}
\affiliation{Brookhaven National Laboratory, Upton, New York 11973, USA}
\affiliation{University of California, Berkeley, California 94720, USA}
\affiliation{University of California, Davis, California 95616, USA}
\affiliation{University of California, Los Angeles, California 90095, USA}
\affiliation{Universidade Estadual de Campinas, Sao Paulo 13131, Brazil}
\affiliation{Central China Normal University (HZNU), Wuhan 430079, China}
\affiliation{University of Illinois at Chicago, Chicago, Illinois 60607, USA}
\affiliation{Creighton University, Omaha, Nebraska 68178, USA}
\affiliation{Czech Technical University in Prague, FNSPE, Prague, 115 19, Czech Republic}
\affiliation{Nuclear Physics Institute AS CR, 250 68 \v{R}e\v{z}/Prague, Czech Republic}
\affiliation{Frankfurt Institute for Advanced Studies FIAS, Frankfurt 60438, Germany}
\affiliation{Institute of Physics, Bhubaneswar 751005, India}
\affiliation{Indian Institute of Technology, Mumbai 400076, India}
\affiliation{Indiana University, Bloomington, Indiana 47408, USA}
\affiliation{Alikhanov Institute for Theoretical and Experimental Physics, Moscow 117218, Russia}
\affiliation{University of Jammu, Jammu 180001, India}
\affiliation{Joint Institute for Nuclear Research, Dubna, 141 980, Russia}
\affiliation{Kent State University, Kent, Ohio 44242, USA}
\affiliation{University of Kentucky, Lexington, Kentucky, 40506-0055, USA}
\affiliation{Korea Institute of Science and Technology Information, Daejeon 305-701, Korea}
\affiliation{Institute of Modern Physics, Lanzhou 730000, China}
\affiliation{Lawrence Berkeley National Laboratory, Berkeley, California 94720, USA}
\affiliation{Massachusetts Institute of Technology, Cambridge, Massachusetts 02139-4307, USA}
\affiliation{Max-Planck-Institut fur Physik, Munich 80805, Germany}
\affiliation{Michigan State University, East Lansing, Michigan 48824, USA}
\affiliation{Moscow Engineering Physics Institute, Moscow 115409, Russia}
\affiliation{National Institute of Science Education and Research, Bhubaneswar 751005, India}
\affiliation{Ohio State University, Columbus, Ohio 43210, USA}
\affiliation{Institute of Nuclear Physics PAN, Cracow 31-342, Poland}
\affiliation{Panjab University, Chandigarh 160014, India}
\affiliation{Pennsylvania State University, University Park, Pennsylvania 16802, USA}
\affiliation{Institute of High Energy Physics, Protvino 142281, Russia}
\affiliation{Purdue University, West Lafayette, Indiana 47907, USA}
\affiliation{Pusan National University, Pusan 609735, Republic of Korea}
\affiliation{University of Rajasthan, Jaipur 302004, India}
\affiliation{Rice University, Houston, Texas 77251, USA}
\affiliation{University of Science and Technology of China, Hefei 230026, China}
\affiliation{Shandong University, Jinan, Shandong 250100, China}
\affiliation{Shanghai Institute of Applied Physics, Shanghai 201800, China}
\affiliation{Temple University, Philadelphia, Pennsylvania 19122, USA}
\affiliation{Texas A\&M University, College Station, Texas 77843, USA}
\affiliation{University of Texas, Austin, Texas 78712, USA}
\affiliation{University of Houston, Houston, Texas 77204, USA}
\affiliation{Tsinghua University, Beijing 100084, China}
\affiliation{United States Naval Academy, Annapolis, Maryland, 21402, USA}
\affiliation{Valparaiso University, Valparaiso, Indiana 46383, USA}
\affiliation{Variable Energy Cyclotron Centre, Kolkata 700064, India}
\affiliation{Warsaw University of Technology, Warsaw 00-661, Poland}
\affiliation{Wayne State University, Detroit, Michigan 48201, USA}
\affiliation{World Laboratory for Cosmology and Particle Physics (WLCAPP), Cairo 11571, Egypt}
\affiliation{Yale University, New Haven, Connecticut 06520, USA}
\affiliation{University of Zagreb, Zagreb, HR-10002, Croatia}

\author{L.~Adamczyk}\affiliation{AGH University of Science and Technology, Cracow 30-059, Poland}
\author{J.~K.~Adkins}\affiliation{University of Kentucky, Lexington, Kentucky, 40506-0055, USA}
\author{G.~Agakishiev}\affiliation{Joint Institute for Nuclear Research, Dubna, 141 980, Russia}
\author{M.~M.~Aggarwal}\affiliation{Panjab University, Chandigarh 160014, India}
\author{Z.~Ahammed}\affiliation{Variable Energy Cyclotron Centre, Kolkata 700064, India}
\author{I.~Alekseev}\affiliation{Alikhanov Institute for Theoretical and Experimental Physics, Moscow 117218, Russia}
\author{J.~Alford}\affiliation{Kent State University, Kent, Ohio 44242, USA}
\author{A.~Aparin}\affiliation{Joint Institute for Nuclear Research, Dubna, 141 980, Russia}
\author{D.~Arkhipkin}\affiliation{Brookhaven National Laboratory, Upton, New York 11973, USA}
\author{E.~C.~Aschenauer}\affiliation{Brookhaven National Laboratory, Upton, New York 11973, USA}
% ...

\collaboration{STAR Collaboration}\noaffiliation

\date{\today}% It is always \today, today,
%\linenumbers

\begin{abstract}
Balance functions have been measured in terms of relative pseudorapidity ($\Delta \eta$) for charged particle pairs at the Relativistic Heavy-Ion Collider (RHIC) from Au+Au collisions at $\sqrt{s_{\rm NN}}$ = 7.7 GeV to 200 GeV using the STAR detector.  These results are compared with balance functions measured at the Large Hadron Collider (LHC) from Pb+Pb collisions at $\sqrt{s_{\rm NN}}$ = 2.76 TeV by the ALICE Collaboration.  The width of the balance function decreases as the collisions become more central and as the beam energy is increased.  In contrast, the widths of the balance functions calculated using shuffled events show little dependence on centrality or beam energy and are larger than the observed widths.  Balance function widths calculated using events generated by UrQMD are wider than the measured widths in central collisions and show little centrality dependence.  The measured widths of the balance functions in central collisions are consistent with the delayed hadronization of a deconfined quark gluon plasma (QGP).  The narrowing of the balance function in central collisions at $\sqrt{s_{\rm NN}}$ = 7.7 GeV implies that a QGP is still being created at this relatively low energy.
\end{abstract}

\maketitle

Event-by-event charge correlations and fluctuations can be used as a tool to study the dynamics of hadronization in relativistic heavy-ion collisions \cite{stephanov_fluc_tricritical, stephanov_fluc_qcd_crit, fluc_collective, signatures, charge_fluct, charge_fluct2, fluctuations_review_heiselberg, fluct3, fluct4, fluct5, methods_fluctuations, stephanov_thermal_fluc_pion, hijing_jet_study, gavin_pt_fluc, ceres_pt, wa98_fluc, na49_fluc, star_deta_dphi_cf_200, star_deta_dphi_cf, star_pt_fluc_excitation, star_pt_fluc, star_charge_fluc, star_balance, phenix_net_charge_fluc, phenix_pt_fluc, phenix_pt_2004, balance_theory, balance_distortions_jeon, balance_distortions, balance_blastwave, kpi_fluctuations, net_protons, net_charge}. One such observable, the balance function \cite{balance_theory, balance_distortions_jeon, balance_distortions, balance_blastwave}, is sensitive to the correlation of balancing charges. The basic idea of the balance function is that charge is created in balancing pairs that originate from the same point in space and time.  By means of a like-sign subtraction, the balance function can yield the distribution of relative momentum between the balancing charges. Balance functions are sensitive to the mechanisms of charge formation and the subsequent relative diffusion of the balancing charges \cite{balance_theory} and are also affected by the freeze-out temperature and radial flow \cite{balance_distortions_jeon}. Model calculations show that collective flow is not sufficient to explain the balance-function widths measured in central Au+Au collisions at $\sqrt{s_{\rm NN}}$ = 200 GeV \cite{balance_distortions_jeon,SchlictingPrattPRC,balance_blastwave}. Balance functions for central collisions have been shown to be consistent with blast-wave models where the balancing charges are required to come from regions with similar collective flow \cite{balance_blastwave}. The inferred high degree of correlation in coordinate space has been postulated as a signal for delayed hadronization \cite{balance_theory}.  In central collisions, a deconfined system of quarks and gluons is created, which cools and expands \cite{STAR_White_Paper}.  Most of the observed balancing charges are then created when the deconfined system hadronizes, which limits the time available for the balancing charges to diffuse away from one another.  This leads to tighter correlations in coordinate space of balancing charges, and due to collective motion, results in tighter correlations in relative momentum and relative rapidity.  Alternatively, if the charges are created early (on the order of 1 fm/$c$), the balancing charges are less correlated in the final state because the balancing charges have more time to move apart from one another.  Thus, a narrow balance function in terms of relative pseudorapidity or relative rapidity in central collisions compared with peripheral collisions implies delayed hadronization.

The balance function is a conditional distribution \cite{balance_theory}, which can be written as:
\begin{eqnarray}
B(\Delta \eta ) &=& {\frac{1}{2}} 
\left\{
{\frac{N _{+-}(\Delta \eta )  - N _{++}(\Delta \eta  )}{N_+}} \right. \nonumber \\ & & \left.
+ {\frac{N_{-+}(\Delta \eta  ) -   N_{--}(\Delta \eta )}{N_-}} \right\}.
\label{balance_function}
\end{eqnarray}

\noindent
The balance function in terms of $\Delta \eta$ ($B(\Delta \eta)$) represents the probability of seeing a particle which has a relative pseudorapidity $\Delta \eta$ with respect to its opposite sign partner, given the condition that its opposite sign partner has already been seen inside the detector. Specifically, $N _{+-}(\Delta \eta) $ is the histogram of $\Delta \eta$ for all the positive-negative charged particle pairs in an event.  It is then accumulated over all events.  Similarly, the histograms for $N_{++}$, $N_{--}$, and $N_{-+}$ are calculated. For the denominators, $N_{+(-)}$ is the number of positive (negative) particles integrated over all events.

\begin{table}
\begin{center}
  \begin{tabular}{c c c}
    \hline
    \hline
   $\sqrt{s_{\rm NN}}$ (GeV) & Year & Events (M) \\ 
   \hline
   200 & 2010 & 32 \\
   62.4 & 2010 & 15 \\
   39 & 2010 & 10 \\
   27 & 2011 & 28 \\
   19.6 & 2011 & 15 \\
   11.5 & 2010 & 7.7 \\
   7.7 & 2010 & 2.2 \\
   \hline
   \hline
\end{tabular}
\caption{\label{tab:tab1_1_01} Summary of the data used in this analysis.}
\end{center}
\end{table}

The system size and centrality dependence of the balance function for all charged particles has been
studied by the NA49 Collaboration at $\sqrt{s_{\rm NN}}$ = 17.3 GeV for $p$+$p$, C+C, Si+Si, and Pb+Pb collisions \cite{NA49_PRC_2005}. The balance function for all charged particles narrows in central Pb+Pb collisions at 17.3 GeV and the widths of the balance functions for $p$+$p$, C+C, Si+Si, and Pb+Pb collisions scale with the number of participating nucleons. The NA49 Collaboration has also published results \cite{NA49_PRC_2007} for the rapidity dependence and beam energy dependence of the balance function for all charged particles for Pb+Pb collisions from $\sqrt{s_{\rm NN}}$ = 6.3 GeV to 17.3 GeV. The balance function was observed to narrow in central collisions for midrapidity, but did not narrow at forward rapidity.  The authors of Ref. \cite{NA49_PRC_2007} showed that the narrowing of the balance function in terms of $\Delta \eta$ in central collisions was explained with the AMPT (a multiphase transport) model \cite{AMPT} incorporating delayed hadronization, while models such as HIJING (heavy-ion jet interaction generator, version 1.38, default parameters) \cite{HIJING} and UrQMD (ultra relativistic quantum molecular dynamics, version 3.3, with default parameters)  \cite{UrQMD} failed to reproduce the observed narrowing.

The STAR Collaboration has presented a study of the longitudinal scaling of the balance function in Au+Au collisions at $\sqrt{s_{\rm NN}}$ = 200 GeV \cite{STAR_long_scaling}.  STAR has published results for balance functions from $p$+$p$, $d$+Au, and Au+Au collisions at $\sqrt{s_{\rm NN}}$ =  130 and 200 GeV in terms of $\Delta \eta$, relative rapidity ($\Delta y$), relative azimuthal angle ($\Delta \phi$), and invariant relative momentum ($q_{\rm inv}$) for all charged particles, for charged pions, and for charged kaons \cite{STAR_PRL,STAR_PRC}. The balance functions for all charged particles and for charged pions narrow as the events become more central while balance functions calculated using HIJING and UrQMD showed no centrality dependence. The ALICE Collaboration has recently published measurements \cite{ALICE_Balance_PLB} from Pb+Pb collisions at $\sqrt{s_{\rm NN}}$ = 2.76 TeV that also show that the balance functions in terms of $\Delta \eta$ narrow in central collisions.

\begin{figure}
\includegraphics[width=3.25in]{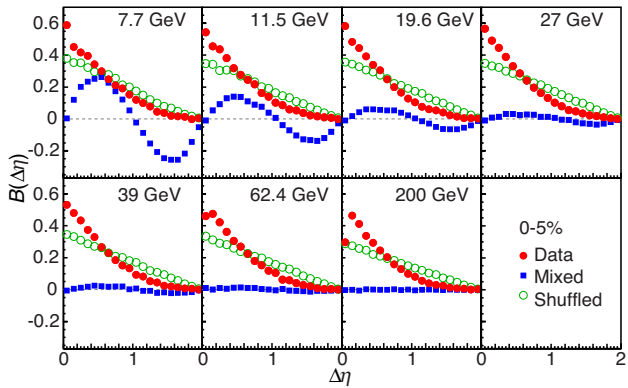}
\caption{\label{fig:fig01}(Color online)  The balance function in terms of $\Delta \eta$ for all charged particles with $0.2 < p_{\rm T} < 2.0$ GeV/$c$ from central Au+Au collisions (0-5\%) for $\sqrt{s_{\rm NN}}$ from 7.7 to 200 GeV.  The data are the measured balance functions corrected by subtracting balance functions calculated using mixed events.  Also shown are balance functions calculated using shuffled events. }
\end{figure}

In this paper, we report measurements of balance functions for all charged particles with $0.2 < p_{\rm T} < 2.0$ GeV/$c$ in terms of relative pseudorapidity ($\Delta \eta$) from Au+Au collisions at $\sqrt{s_{\rm NN}}$ = 7.7, 11.5, 19.6, 27, 39, 62.4, and 200 GeV.  We observe that the balance functions narrow in central collisions and narrow as the beam energy is increased.  We compare with the results from ALICE for Pb+Pb collisions at $\sqrt{s_{\rm NN}}$ = 2.76 TeV by restricting STAR's acceptance to $\Delta \eta \le 1.6$ to match the acceptance of ALICE, correcting for the acceptance of STAR in $\eta$, and calculating the width of the balance function over the range $0.1 < \Delta \eta < 1.6$, where the lower limit of 0.1 is chosen to suppress effects from interpair correlations [e.g., Hanbury-Brown and Twiss (HBT) and final-state interactions].  We observe that the balance function in terms of $\Delta \eta$ narrows as the beam energy is raised from $\sqrt{s_{\rm NN}}$ = 7.7 GeV to 2.76 TeV.   When the observed balance function widths are scaled by the width observed in the most peripheral bin, the relative widths still decrease as the events become more central and as the beam energy is increased.  These results contrast with those presented in Ref. \cite{ALICE_Balance_PLB}, where the scaled balance function widths are shown to be nearly the same at RHIC and LHC energies.  The present observations are consistent with the concept of delayed hadronization of a deconfined quark gluon plasma (QGP) with the deconfined system having a longer lifetime at the highest energy.  The narrowing of the balance function in central collisions at $\sqrt{s_{\rm NN}}$ = 7.7 GeV implies that a QGP might still be created at this relatively low energy.

\begin{figure}
\includegraphics[width=3.25in]{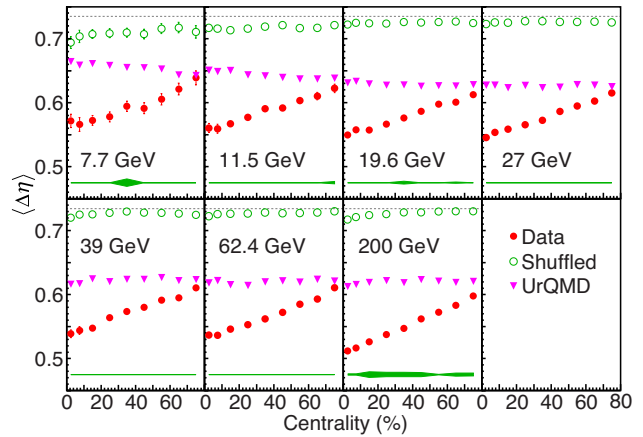}
\caption{\label{fig:fig02}(Color online) Energy dependence of the balance function widths compared with the widths of the balance functions calculated using shuffled events.  Also shown are the balance function widths calculated using UrQMD.  The dashed line represents the width of the balance function calculated using shuffled events for a constant $dN/d\eta$ distribution.  Error bars represent the statistical error and the shaded bands represent the systematic error.}
\end{figure}

The data were taken with the STAR detector \cite{TPCref} during the years 2010 and 2011. Table~\ref{tab:tab1_1_01} shows a summary of the data sets used in this analysis. Au+Au collisions were studied at seven beam energies ranging from 7.7 GeV to 200 GeV. The centrality of each collision was determined according to the measured charged hadron multiplicity within the pseudorapidity range $|\eta| < 0.5$. Nine centrality bins were used: 0-5\% (most central), 5-10\%, 10-20\%, 20-30\%, 30-40\%, 40-50\%, 50-60\%, 60-70\%, and 70-80\% (most peripheral).  At each of the seven beam energies, the average number of participating nucleons, $N_{\rm part}$, is calculated for each of the nine centrality bins using a Glauber model.  To ensure a more uniform detector acceptance, events were accepted only when the position of the reconstructed primary vertex was within 30 cm of the center of STAR ($\left| {{z_{{\rm{vertex}}}}} \right| < 30$ cm). In addition, the radial position of the primary vertex was required to be less than 2 cm from the center of the beam line to avoid beam pipe events. All events were required to have at least one matched track with the STAR Time-of-Flight (TOF) system \cite{TOF} to suppress pile-up events.

\begin{figure}
\includegraphics[width=3.25in]{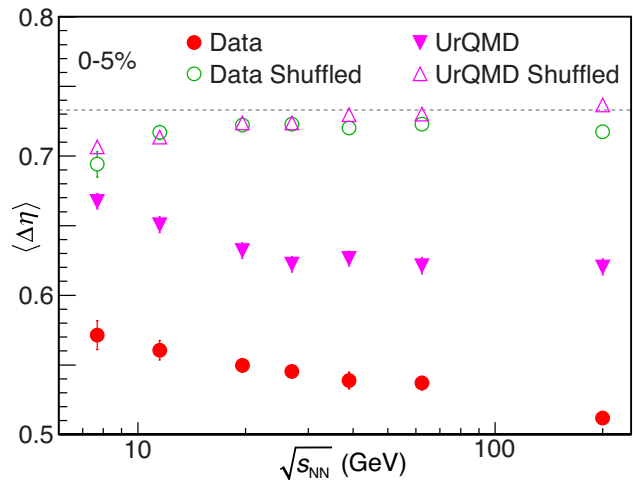}
\caption{\label{fig:fig03}(Color online) Balance function widths for the most central events (0-5\%) compared with balance function widths calculated using shuffled events.  Also shown are balance function widths calculated using UrQMD and shuffled UrQMD events.  The dashed line represents the width of the balance function calculated using shuffled events for a constant $dN/d\eta$ distribution.}
\end{figure}

All tracks in the Time Projection Chamber (TPC) were required to have more than 15 measured space points along the trajectory. The ratio of the number of reconstructed space points to possible space points along the track was required to be greater than 0.52 to avoid  track splitting. Tracks in the TPC were characterized by the distance of closest approach (DCA), which is the smallest distance between the projection of the track and the measured event vertex. To suppress decay effects and background, all tracks were required to have a DCA less than 3 cm. A transverse momentum cut of $0.2 < p_{\rm T} < 2.0$ GeV/$c$ and a pseudorapidity cut of $| \eta | < 1.0$ were applied.

\begin{figure}
\includegraphics[width=3.25in]{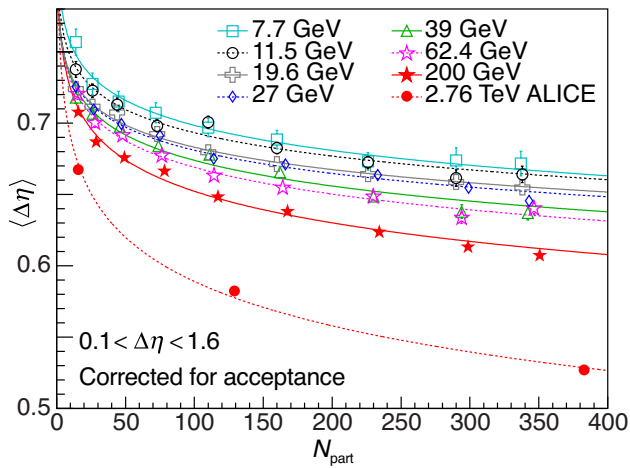}
\caption{\label{fig:fig04}(Color online) Acceptance-corrected balance function widths for Au+Au measured over the range $0.1 < \Delta \eta < 1.6$ compared with similar results from Pb+Pb collisions from ALICE \cite{ALICE_Balance_PLB}.  Only statistical errors are shown.  Lines represent fits of the form $a + b(N_{\rm part})^{0.01}$.}
\end{figure}

In addition to real data, mixed events and shuffled events were also used in this analysis.  Mixed events are created by grouping the events according to bins in centrality and bins in the position of the reconstructed vertex of the event along the beam direction. Ten centrality bins and five bins in $z_{\rm vertex}$ were used.   A set of mixed events is created by taking one track chosen at random from an event, which is selected according to the bin in centrality and the bin in event vertex position. A mixed event includes no more than one track from any observed event. This mixed-event data set has the same number of events with the same multiplicity distribution as the original data set but all correlations are removed. The mixed-event subtraction was important, especially at low energies, to account for the effects caused by unbalanced positive charges in each event.

Shuffled events are produced by randomly shuffling the charges of the particles in each event, which removes the charge correlations while retaining global charge conservation. Because shuffling uniformly distributes a particle's balancing partner across the measured phase space, balance functions calculated using shuffled events can be used to gauge the widest balance functions that one can measure within the experimental acceptance of STAR.

Fig.~\ref{fig:fig01} shows the balance functions in terms of $\Delta \eta$ for all charged particles for Au+Au collisions at $\sqrt{s_{\rm NN}}$ = 7.7, 11.5, 19.6, 27, 39, 62.4, and 200 GeV for the most central events (0-5\%) along with balance functions calculated using shuffled events and balance functions calculated using mixed events. The data shown in the figure are the measured balance functions corrected by subtracting balance functions calculated using mixed events. These data have not been corrected for efficiency or acceptance.  The conclusions of this paper involve the width of the balance function in which the efficiency cancels out.  The model calculations shown in this paper use the STAR acceptance.  When comparisons are made with the width of the balance functions reported by ALICE \cite{ALICE_Balance_PLB}, the STAR data are corrected for acceptance.

At the lower energies, the balance functions calculated using mixed events exhibit an oscillatory distribution that is 0 at $\Delta\eta$ = 0, has a positive value at $\Delta\eta$ = 0.5, is 0 at $\Delta\eta$ = 1, has a negative value at $\Delta\eta$ = 1.5 and is 0 again at $\Delta\eta$ = 2.  This oscillatory behavior lessens as the events become more peripheral and as the beam energy is increased.  This effect is due to unbalanced positive charge that is not subtracted by the same sign subtraction inherent in the balance function.  The additional positive charges are dominantly protons and have a different $dN/d\eta$ distribution than the negative charges that are dominantly pions.  The $dN/d\eta$ distributions for the difference between the positive and negative charges have minima at $\eta$ = -1, $\eta$ = 0, and $\eta$ = 1.  Thus, when the balance function in terms of $\Delta\eta$ is calculated for mixed events at the lower energies and in more central collisions, the oscillatory distribution is obtained.  At $\sqrt{s_{\rm NN}}$ = 200 GeV, the balance functions calculated using mixed events are zero for all centralities, which indicates that the amount of unbalanced positive charge is small.  As the beam energy is decreased, the unbalanced positive charge increases and the balance functions calculated using mixed events become significant.

The corrected balance functions are narrower than the balance functions calculated using shuffled events and the balance functions narrow as the events become more central (see below).  Also visible are the effects of interpair correlations [HBT and final-state interactions] that model calculations have shown to be significant for $\Delta \eta \lesssim 0.1$ \cite{balance_distortions}.  Specifically, $B\left( {\Delta \eta } \right)$ for $\Delta \eta  < 0.1$ is noticably higher than the trend of the remaining points at 7.7 GeV while $B\left( {\Delta \eta } \right)$ for $\left\langle {\Delta \eta } \right\rangle  < 0.1$ is lower than the trend at 200 GeV.
The width of the balance function is characterized in terms of a weighted average:
\begin{eqnarray}
\label{WA}
\left\langle {\Delta \eta } \right\rangle  = \frac{{\sum\limits_{i = {i_{{\rm{lower}}}}}^{{i_{{\rm{upper}}}}} {B\left( {\Delta {\eta _i}} \right)\Delta {\eta _i}} }}{{\sum\limits_{i = {i_{{\rm{lower}}}}}^{{i_{{\rm{upper}}}}} {B\left( {\Delta {\eta _i}} \right)} }}.
\end{eqnarray}
Here $i$ is the bin number and $B( {\Delta \eta _i )}$ is the value of the balance function for the relative pseudorapidity bin $\Delta \eta_i$.  The weighted average is calculated over a range in $\Delta \eta$ chosen to minimize contributions from HBT and Coulomb effects ($\Delta \eta \ge 0.1$) and maximize the acceptance of STAR ($\Delta \eta \le 2.0$).

Fig.~\ref{fig:fig02} shows the balance function widths for Au+Au collisions from $\sqrt{s_{\rm NN}}$ = 7.7 GeV to 200 GeV for nine centrality bins.  The widths are calculated for $0.1 < \Delta \eta  < 2.0$ to remove the distortions caused by interpair correlations for $\Delta \eta < 0.1 $ \cite{balance_distortions}.  The widths of the balance functions calculated using shuffled events are larger than the widths of the balance functions calculated using data.  The widths of the balance functions using shuffled events shown in Fig. \ref{fig:fig02} are close to the value 0.733, which one would expect for shuffled events from a flat $dN/d\eta$ distribution over the range $-1 < \eta < 1$.  The data show a smoothly decreasing width with increasing beam energy and as the collisions become more central.  Fig.~\ref{fig:fig02} also shows the widths of balance functions calculated using UrQMD.  The UrQMD calculations are analyzed in the same way as the data with the balance functions calculated from mixed UrQMD events being subtracted from the balance functions calculated using UrQMD. For beam energies below 20 GeV, the balance function widths from UrQMD increase as events become more central whereas the measured widths decrease.  Above 20 GeV, the balance function widths from UrQMD show little centrality dependence.  In peripheral collisions, the balance functions widths from UrQMD approach the value of the measured balance function widths. The UrQMD model is a hadronic model that does not have a deconfined phase and has little flow.  The early hadronization time of the particles calculated using UrQMD combined with the strong interaction between final state particles causes the larger balance function widths in central collisions while the balance function widths calculated using UrQMD are close to the measured balance function widths in peripheral collisions. 

One source of systematic errors was estimated by studying the difference between the 200 GeV results from three different runs (in 2007, 2010, and 2011) that used different tracking software and incorporated different hardware configurations in STAR.  A second source of systematic errors was estimated by varying the DCA used to select tracks.  A third source of systematic error was estimated by varying the range of the $z_{\rm vertex}$ of events accepted in STAR.  The systematic errors in the extracted widths are shown as a shaded band in Fig.~\ref{fig:fig02}.  Note that the systematic error in the width for the most central bin at all energies was of the same order or less than the statistical errors.

Fig.~\ref{fig:fig03} shows the width of the balance function in terms of $\Delta \eta$ for central collisions (0-5\%) as a function of beam energy.  The measured balance function widths decrease smoothly with increasing beam energy.  Also shown are the widths of the balance function calculated using events generated with the UrQMD model.   Although the energy trends for the width of the balance function in UrQMD and data appear similar, the data are much narrower, and as shown in Fig. 2, UrQMD predicts the wrong centrality dependence.  The widths of the balance function calculated from shuffled events from both the data and UrQMD are much larger than the widths calculated using the data.  The decrease of the shuffled widths at the lower beam energies reflects the fact that the $dN/d\eta$ distributions at the lower beam energies are not completely flat.  The fact that the measured balance function widths decrease smoothly with increasing beam energy and are much smaller than the widths predicted by UrQMD is consistent with the idea of delayed hadronization.

\begin{figure}
\includegraphics[width=3.25in]{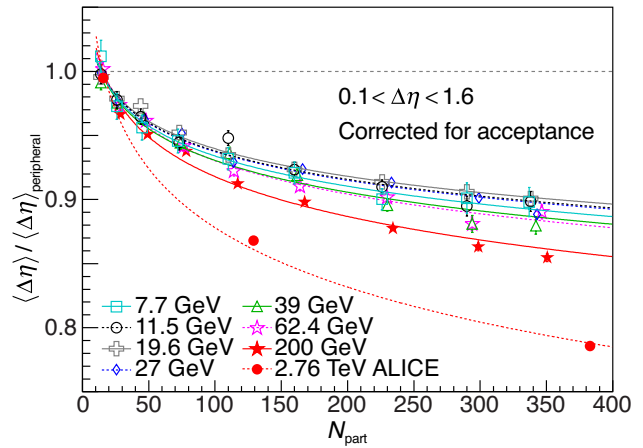}
\caption{\label{fig:fig05}(Color online) Acceptance-corrected balance function widths for Au+Au measured over the range $0.1 < \Delta \eta < 1.6$ normalized to the most peripheral centrality bin compared with similar results from Pb+Pb collisions from ALICE \cite{ALICE_Balance_PLB}.  Only statistical errors are shown. Lines represent fits of the form $a + b(N_{\rm part})^{0.01}$.}
\end{figure}
 
To compare with the balance functions measured by ALICE, we correct our measured balance functions for the acceptance of STAR in $\eta$ using the expression \cite{balance_distortions_jeon}:
\begin{eqnarray}
\label{BalanceEtaCorrection}
{B_{\Delta {\eta _{{\rm{max}}}} = 2}}\left( {\Delta \eta } \right) = {B_\infty }\left( {\Delta \eta } \right)\left( {1 - \frac{{\Delta \eta }}{2}} \right),
\end{eqnarray}
where ${B_\infty }\left( {\Delta \eta } \right)$ is the STAR balance function corrected for acceptance in $\eta$ assuming that STAR's acceptance is constant for $-1 < \eta < 1$, and ${B_{\Delta {\eta _{{\rm{max}}}} = 2}}$ is the measured STAR balance function that is not corrected for acceptance in $\eta$.
For the comparison with the results from ALICE \cite{ALICE_Balance_PLB}, we calculate the widths of the acceptance-corrected balance functions over the range $0.1 \le \Delta \eta \le 1.6$ to suppress effects from interpair correlations and to match the acceptance of ALICE in $\eta$.  Fig.~\ref{fig:fig04} shows these widths as a function of centrality and beam energy for Au+Au collisions.  In the same figure we show the width of the balance function from Pb+Pb collisions at $\sqrt{s_{\rm NN}}$ = 2.76 GeV calculated for three published centralities \cite{ALICE_Balance_PLB} over the same range in $\Delta \eta$.  The balance functions narrow as the beam energy is raised from $\sqrt{s_{\rm NN}}$ = 7.7 GeV up to 2.76  TeV and the balance functions narrow as the collisions become more central.  These observations are consistent with the concept of delayed hadronization.

The authors of Ref. \cite{ALICE_Balance_PLB} assert that the relative decrease of $\Delta \eta$ with centrality does not change appreciably with beam energy.  To address this point, we calculate the ratio of the width of the balance function at each centrality to the width of the balance function in the most peripheral bin at each beam energy, $\left\langle \Delta\eta  \right\rangle /{\left\langle \Delta\eta  \right\rangle _{{\rm{peripheral}}}}$.  Because the peripheral bin at the lower energies has low statistics, we first fit the measured widths at each beam energy with a function of the form $a + b(N_{\rm part})^{0.01}$ and then take the ratio of the measured widths to the width of the fitted distribution at the most peripheral centrality bin.  These results are shown in Fig.~\ref{fig:fig05}.

The relative decrease of the balance function width is much larger at $\sqrt{s_{\rm NN}}$ = 2.76 TeV.  The relative decrease then gets smaller as the beam energy is lowered.  Thus, we observe that the relative decrease of the balance function width clearly changes with beam energy.  The difference between the present analysis and the one presented by the authors of Ref. \cite{ALICE_Balance_PLB} is that we calculate the widths over the range $0.1 \le \Delta \eta \le 1.6$ for both experiments, which minimizes the contributions from interpair correlations and maximizes the measured acceptance. In contrast, the authors of Ref. \cite{ALICE_Balance_PLB} calculated the widths over the range $0.0 \le \Delta \eta \le 1.6$ for the ALICE balance functions and $0.0 \le \Delta \eta \le 2.0$ for the STAR balance functions. We do not compare with the results from NA49 \cite{NA49_PRC_2007} here because the acceptance of NA49 in $\eta$ is relatively small.    

Model calculations \cite{balance_distortions_jeon,balance_distortions} show that a part of the narrowing of the balance function in central collisions is due to radial flow.  Thus, the fact that the balance functions in central Pb+Pb collisions at 2.76 TeV are narrower than those in central Au+Au collisions at 200 GeV may be due to an increase in radial flow.  One would expect that the balance function in terms of $\Delta \eta$ would be narrower for a longer-lived deconfined QGP, which implies that these results are consistent with the concept of delayed hadronization.

In conclusion, we observe that the balance function in terms of $\Delta \eta$ is narrow in central collisions of Au+Au.  At higher beam energies, the balance function in terms of $\Delta \eta$ in central collisions of Pb+Pb is even narrower.  This observed narrowing is consistent with the concept of the delayed hadronization of a deconfined QGP produced in these collisions.  We observe that the balance functions in Au+Au events at $\sqrt{s_{\rm NN}}$ = 7.7 GeV still narrow as the collisions become more central, which suggests that a deconfined QGP might still be produced at this relatively low beam energy.

We thank the RHIC Operations Group and RCF at BNL, the NERSC Center at LBNL, the KISTI Center in Korea, and the Open Science Grid consortium for providing resources and support. This work was supported in part by the Offices of NP and HEP within the U.S. DOE Office of Science, the U.S. NSF, CNRS/IN2P3, FAPESP CNPq of Brazil,  the Ministry of Education and Science of the Russian Federation, NNSFC, CAS, MoST and MoE of China, the Korean Research Foundation, GA and MSMT of the Czech Republic, FIAS of Germany, DAE, DST, and CSIR of India, the National Science Centre of Poland, National Research Foundation (NRF-2012004024), the Ministry of Science, Education and Sports of the Republic of Croatia, and RosAtom of Russia.

\end{document}